\documentclass[conference]{IEEEtran}
\IEEEoverridecommandlockouts
\usepackage{cite}
\usepackage{amsmath,amssymb,amsfonts}
\usepackage{algorithmic}
\usepackage{graphicx}
\usepackage{textcomp}
\usepackage{xcolor}
\def\BibTeX{{\rm B\kern-.05em{\sc i\kern-.025em b}\kern-.08em
    T\kern-.1667em\lower.7ex\hbox{E}\kern-.125emX}}
    
\usepackage{enumitem}
\usepackage{booktabs}
\usepackage{microtype}

\usepackage{tabularx}

\newcommand{\compresslist}{
  \setlength{\itemsep}{1pt}
  \setlength{\parskip}{0pt}
  \setlength{\parsep}{0pt}
}

\newcommand{\comm}[1]{}

\usepackage[T1]{fontenc}

\begin{document}

\title{Effects of Task Type and Wall Appearance on Collision Behavior in Virtual Environments
\thanks{978-1-6654-3886-5/21/\$31.00 ©2021 IEEE}}

\author{\IEEEauthorblockN{1\textsuperscript{st} Sebastian Cmentowski}
\IEEEauthorblockA{\textit{High Performance Computing Group} \\
\textit{University of Duisburg-Essen}\\
Duisburg, Germany \\
sebastian.cmentowski@uni-due.de}
\and
\IEEEauthorblockN{2\textsuperscript{nd} Jens Kr\"uger}
\IEEEauthorblockA{\textit{High Performance Computing Group} \\
\textit{University of Duisburg-Essen}\\
Duisburg, Germany \\
jens.krueger@uni-due.de}
}

\comm{
\author{\IEEEauthorblockN{Authors left anonymous for submission}
\IEEEauthorblockA{\textit{} \\
\textit{}\\
 \\
}
}}

\maketitle

\begin{abstract}
Driven by the games community, virtual reality setups have lately evolved into affordable and consumer-ready mobile headsets. However, despite these promising improvements, it remains challenging to convey immersive and engaging VR games as players are usually limited to experience the virtual world by vision and hearing only. One prominent example of such open challenges is the disparity between the real surroundings and the virtual environment. As virtual obstacles usually do not have a physical counterpart, players might walk through walls enclosing the level. Thus, past research mainly focussed on multisensory collision feedback to deter players from ignoring obstacles. However, the underlying causative reasons for such unwanted behavior have mostly remained unclear.

Our work investigates how task types and wall appearances influence the players' incentives to walk through virtual walls. Therefore, we conducted a user study, confronting the participants with different task motivations and walls of varying opacity and realism. Our evaluation reveals that players generally adhere to realistic behavior, as long as the experience feels interesting and diverse. Furthermore, we found that opaque walls excel in deterring subjects from cutting short, whereas different degrees of realism had no significant influence on walking trajectories. Finally, we use collected player feedback to discuss individual reasons for the observed behavior.
\end{abstract}


\section{Introduction}
Exploring extensive virtual worlds is a challenging task. Despite the development of more than 100 virtual navigation techniques, real walking is still considered the gold standard for VR locomotion~\cite{luca2021locomotion, Usoh1999}. Matching the virtual movement to the physical steps offers precise control and assures a realistic and natural experience. However, using the real movement also introduces additional challenges that must be considered in the development phase. Although current headsets, such as Oculus Quest 2~\cite{oculus}, already handle sufficient room-scale tracking of the head-mounted display (HMD) and the controllers, the disparity between the virtual world and the real surroundings remains a problem~\cite{Simeone2017, razzaque2001redirected}.

These differences between the physical playspace and the VR scenario typically fall into one of two groups: The first are virtual obstacles, such as walls, that do not have a physical counterpart. Most VR setups---except specifically designed lab environments---do not offer matching haptic proxies for the virtual objects~\cite{Simeone2015}. Consequently, players could easily grasp or walk through these immaterial obstacles, which breaks coherency and might spoil the game experience~\cite{Boldt2018}. In the other group fall real obstructions, such as a bench standing in the players' living room, that are not visible to the players but pose an imminent risk of injury~\cite{Simeone2017}. Therefore, current VR systems typically mark the playspace's borders with easily identifiable virtual walls to deter players from leaving this area~\cite{hartmann2019}.

Even though the dangers in the second case are much more critical than potential \textit{breaks-in-presence}~\cite{Slater2000Presence}, both types of discrepancies between the real and virtual worlds share a common aspect: Preventing them requires virtual obstacles that the players do not ignore -- be it out of curiosity or in the attempt to cut short. Past research has approached this issue by developing various types of auditory, visual, and vibrotactile feedback to notify players of virtual collisions and deter them from willfully ignoring walls~\cite{Boldt2018, Blom2010}. 

However, up to this point, very little work has addressed the underlying questions: What causes players to not adhere to the virtual world's rules in the first place? Studies have indicated that players might ignore virtual obstacles under specific circumstances~\cite{Boldt2018, Ogawa2020}, but they have mainly focused on simple setups and repeating tasks, such as walking between checkpoints. Other research has shown that players generally tend to conform to the rules in highly immersive experiences~\cite{Slater2009PlaceIllusion}. Whether the previously observed collisions are a general phenomenon or are caused by individual properties of the virtual scenario remains unclear.

Therefore, we explored how different task types and appearances of the virtual walls influence the players' incentives to cut short and walk right through the obstacles. Specifically, we conducted a mixed study setup to isolate the observed effects. The participants played a VR game with two similar rounds of carrying objects between checkpoints. For the within-subject part, we varied the task motivation: In one round, the participants had to solve a puzzle by placing different objects on the correct checkpoint. The other round did not offer a similar motive but was designed as a dull and repetitive job. We combined this design with an additional between-subject part: Participants were split into four groups, each being confronted with another wall type, differing in the degree of opacity and the degree of realism.

Our results indicate that the given task type has the greatest influence on player behavior. Most participants ignored the walls only in the repetitive round to finish their task faster. In the more diverse and interesting puzzle level, very few subjects collided with a single wall. We conclude that players mostly prefer to stick to realistic behavior and only deviate if getting bored.  Furthermore, our experiment reveals that opaque surfaces are highly efficient in deterring players from non-adherent behavior as they feel discouraged from not being able to see behind the wall before walking through. Lastly, our different wall designs had a significant impact on the perceived presence. However, this effect did not influence the players' walking behavior as expected. Apart from testing these three potential factors on collisions in virtual environments, we collected aural feedback from participants to discuss individual reasons for the observed behavior.

\section{Related Work}
In this section, we summarize the relevant prior research to this work. Therefore, we start by covering the fundamentals of real walking and locomotion in general. Next, we briefly address haptic feedback before discussing the current state of research on virtual collisions and walking trajectories.

\subsection{Walking in Virtual Environments}
While non-VR games tend to rely on joystick controls for locomotion, these are less favorable in VR, as purely virtual continuous motion is a key factor leading to cybersicknness~\cite{Habgood:2017:HLP:3130859.3131437}. Instead, natural walking has emerged as the gold standard among locomotion techniques in virtual environments. Accurately transferring the players' steps into the virtual world not only prevents cybersickness but also feels most realistic and natural~\cite{Ruddle2009}. Furthermore, it results in higher presence levels compared to other locomotion alternatives such as walking-in-place~\cite{Templeman1999, Usoh1999}. However, natural walking is limited by the play area's physical boundaries, making it challenging to achieve larger explorable virtual environments. Thus, an ever-growing body of research has focused on overcoming this limitation, for instance, by developing novel locomotion metaphors~\cite{boletsis2017new} or altering the user's movements unconsciously~\cite{razzaque2001redirected, Peck2009}.

\subsection{Haptics and Surfaces}
The virtuality of an immersive experience becomes most obvious when players interact with the virtual world and its objects. Touching a wall or grasping an item without feeling a haptic resistance does not even feel close to the familiar multisensorial experience in reality. Therefore, research has focussed on providing surface feedback through passive proxies~\cite{insko2001passive}, haptic retargeting~\cite{azmandian2016haptic}, electro-tactile feedback~\cite{Pamungkas2016}, or electrical muscle stimulation~\cite{Lopes2015, Lopes2017}. Another promising approach, which does not require additional hardware, is the concept of simulated surface constraints~\cite{Burns2006}: This technique elicits the impression of resistive virtual objects by simply displacing the virtual hand from its real counterpart.

\subsection{Virtual Collisions}
According to Blom et al.~\cite{Blom2010}, treating virtual collisions consists of two consecutive parts: collision detection~\cite{Jimenez2001} and collision notification. The first aspect, collision detection, is a mostly solved problem in current game engines used for virtual environments. Therefore, this section focuses on the latter problem: Collision feedback not only prevents unwanted penetration of virtual objects, but may also increase the perceived realism~\cite{Blom2013}.

In reality, we usually notice bumping into objects through a haptic response. As haptic reactions are mostly missing in virtual scenarios, research has investigated the effectiveness of a wide variety of other possible feedback channels, including vision, sound, and vibrotactile impulses. Among the first to investigate possible collision behaviors, Jacobson and Lewis~\cite{Jacobson1997} altered the users' movement in the virtual world, e.g., by stopping them upon impact. However, such manipulations do not apply to real walking, where the virtual movement is always bound to the physical steps.

Therefore, Bloomfield and Badler~\cite{Bloomfield2007} examined the use of vibrotactile actuators to convey better collision feedback and found them more effective than purely visual indicators. While they used a shirt-based tactorsuit to convey the collision impressions, other research has achieved comparable effects with different hardware, such as tactile belts~\cite{Ryu2004}. In a similar study, Blom et al.~\cite{Blom2010} compared vibrotactile feedback using their haptic floor with sound- and controller-based responses. Even though auditory notifications scored worse than the floor feedback, Afonso et al.~\cite{Afonso2011} found spatial sounds to be well suited for preceding collision avoidance.

\subsection{Walking Behavior}
A growing body of research has added evidence to the finding that people tend to act realistically in virtual scenarios that conform to reality. This behavior is particularly seen in situations with high \textit{Place Illusion},  \textit{Plausibility Illusion}~\cite{Slater2009PlaceIllusion}, and visual realism~\cite{Slater2009VisualRealism}. These findings also apply to VR movement. Ruddle et al.~\cite{Ruddle2013} found that using real walking for locomotion causes users to walk around virtual objects. The observed obstacle avoidance trajectories generally conform to real-world walking patterns~\cite{Fink2007, Cirio2013}.

Simeone et al.~\cite{Simeone2017} investigated the effects of different ground textures and virtual obstacles on individual movement behavior. The reported findings are closely related to our research focus: Participants generally hesitated cutting short through immaterial but solid virtual objects. For the case of collisions, Boldt et al.~\cite{Boldt2018} presented a multimodal collision feedback approach combining visual, auditory, and vibrotactile feedback that effectively deters players from ignoring virtual walls. Similarly, Ogawa et al.~\cite{Ogawa2020} showed that the body-ownership effect of realistic full-body avatars also discourages users from penetrating walls.

\section{Study Design}
As explained in the previous section, research has already dealt with possible ways to deter players from walking through walls using multisensory collision feedback or virtual avatars. However, under which circumstances players ignore walls in the first place remains to be investigated. Therefore, we conducted a study to address the identified unclear aspects of wall-related behavior.

\subsection{Research Questions and Hypotheses}
Existing studies have mainly relied on strong incentives to cut short by using heavily repetitive and monotonous tasks~\cite{Boldt2018, Ogawa2020}. Therefore, our first goal was to confirm the players' behavior under more engaging circumstances by using an immersive puzzle-scenario. We hypothesize that more varying tasks lead to fewer wall collisions than repetitve and annoying missions.

Further, it remains unclear whether the type of virtual wall influences the players' incentives to cut short. Is this decision connected to the thematic fitting of the wall? Less well-fitting walls potentially reduce the virtual environment's overall authenticity. Since visual realism is one key factor for conforming behavior according to Slater et al.~\cite{Slater2009VisualRealism}, we assume that abstract walls provoke players to walk through them more often.

Another influencing factor might be the wall's degree of opacity. Viewing the target through an obstructing wall could enforce the players' decision to cut short. Also, this characteristic might decrease the \textit{Plausibility Illusion}~\cite{Slater2009PlaceIllusion} and raise the players' impression that the wall is safe to pass through. A similar finding was already reported by Simeone et al.~\cite{Simeone2017}. Thus, we assume that partly transparent walls or obstacles with holes lead to more wall collisions.

\noindent In summary, our three hypotheses are as follows:

\begin{itemize}[leftmargin=*]\compresslist
\item H1: Repetitive and monotonous tasks provoke significantly more participants to walk through virtual walls than diverse tasks.
\item H2: Participants walk significantly more often through abstract walls than through realistic walls.
\item H3: Opaque surfaces deter more participants from ignoring virtual obstacles than partly transparent surfaces.
\end{itemize}

\noindent Apart from these hypotheses, we were also interested in the particular reasons players decide to either ignore walls or follow real-world' rules.

\begin{itemize}[leftmargin=*]\compresslist
\item RQ1: How do players decide whether they pass through virtual walls or walk around them?
\end{itemize}

\begin{figure}[t!]
\centering
\includegraphics[width=\columnwidth]{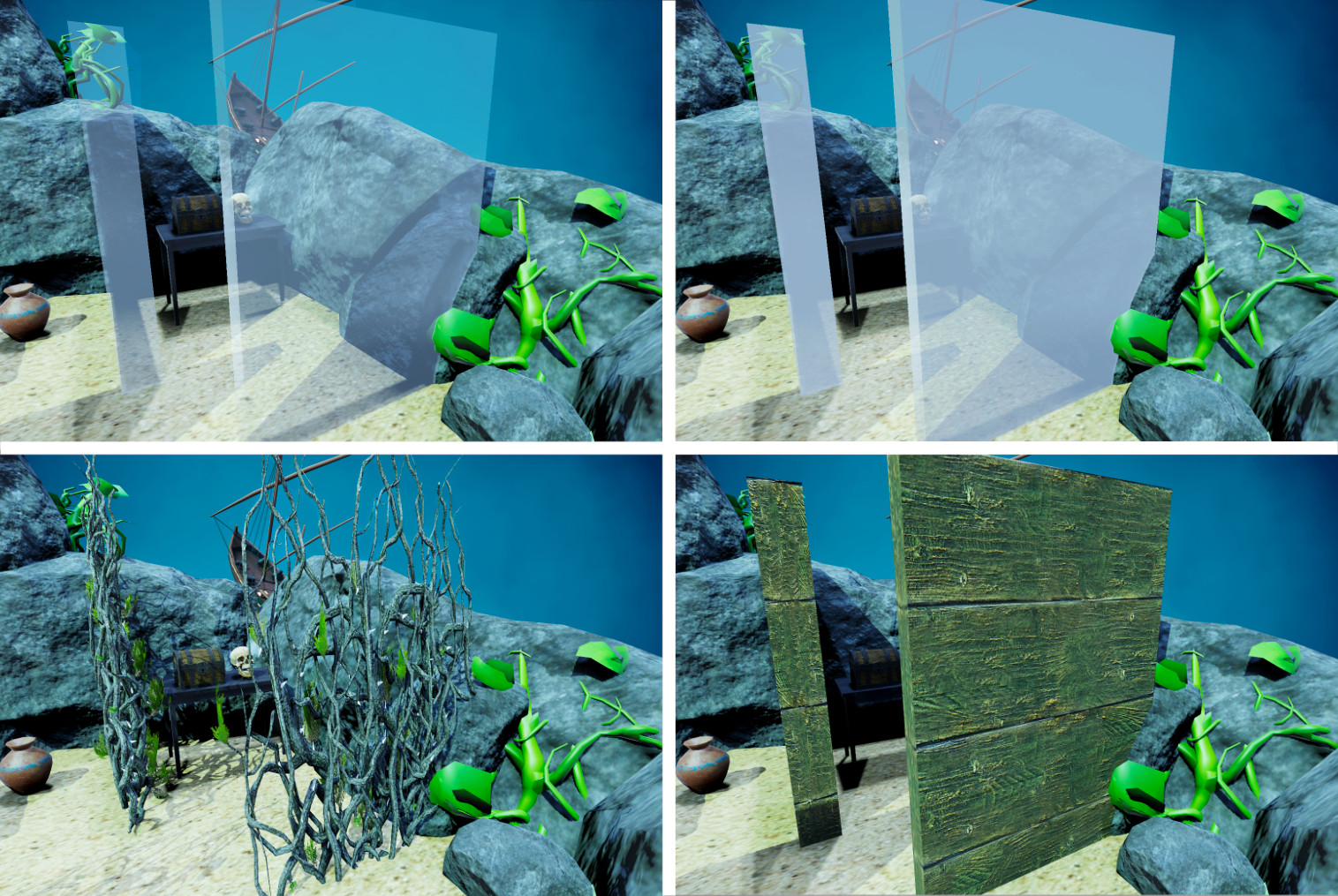}
\caption{Our four tested wall designs: Top row: abstract cuboid wall with 30\% opacity (left) and 60\% opacity (right). Bottom row: twine hedge with holes (left) and opaque wood wall (right), both matching the virtual scenario.}
\label{fig:walldesigns}
\end{figure}

\begin{figure*}[t!]
\centering
\includegraphics[width=2.045\columnwidth]{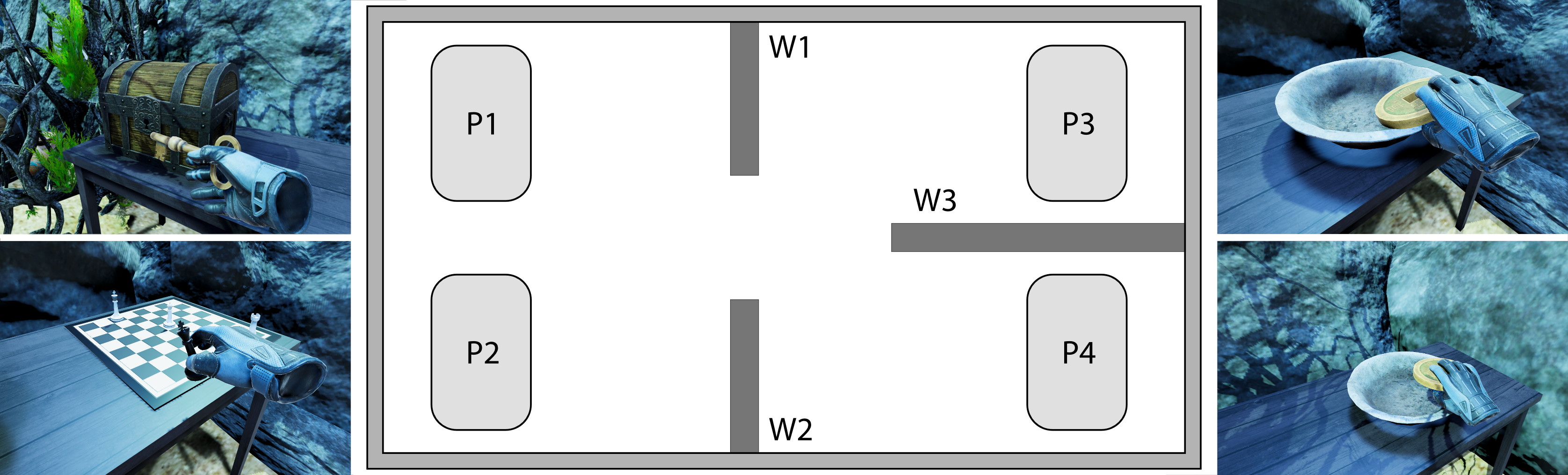}
\caption{Schematic map depicting our testbed environment, featuring four interaction points (P1--P4) and three virtual walls (W1--W3). The images illustrate the different activities: in the puzzle task (left), items and interactions vary. In the repetitive task (right), participants only move coins between bowls.}
\label{fig:taskdesigns}
\end{figure*}

\subsection{Wall Design and Testbed Scenario}
Based on our hypotheses, the virtual walls used in the study should differ in degree of realism and degree of opacity. Considering these requirements, we decided on four wall designs (see Figure~\ref{fig:walldesigns}). Two of these walls are completely abstract blocks with a uniform color. These obstacles differ only in the degree of opacity, with one wall having 30\% and the other 60\% opacity. The other two walls are designed to fit the surrounding scenario thematically. One wall is implemented as a solid wood wall, thus resulting in full opacity. The other wall resembles a hedge consisting of twines. Whereas this design still looks realistic, it offers enough holes to look through it and creates a similar opacity effect as the abstract walls.\vspace{10cm}

\noindent In sum, our four wall designs are:

\begin{itemize}[leftmargin=*]\compresslist
\item A30: Abstract wall design with 30\% opacity
\item A60: Abstract wall design with 60\% opacity
\item RTH: Realistic twine hedge with holes
\item RWW: Realistic and fully opaque wood wall
\end{itemize}

Our surrounding testbed scenario was realized with the Unity game engine~\cite{unity}. The setting is a maritime-themed scenery, featuring boulders and sunken ships (see Figure~\ref{fig:taskdesigns}). Since our research focussed on natural walking, we restricted the virtual environment's size to match our lab, i.e., $16 \, m^2$. This limited area contains four points of interest and three walls, which -- depending on the study condition -- match any of the four styles. The walls separate the interaction points so that the players would have to either walk through them or make a detour through the playspace center. We decided against the fourth wall to avoid cluttering the playspace and producing unintended wall collisions.

\subsection{Tasks}
Testing H1 required two tasks, one diverse and interesting, and the other highly repetitive. Despite these differences, we still aimed for a similar structure in both tasks and only changed the motivation and reasoning behind the required interactions. Our first task is a simple sequential puzzle. The players carry a single item that must be placed in the correct spot to advance with the task and obtain the next object. For instance, the players may use a key to unlock a chest. They are rewarded with a pearl that must be put into an open shell. These subtasks are chained into complex puzzle of adequate length and require the players to constantly walk between the interaction points. 

This task resembles repetitive task designs used in previous studies but adds an engaging motivation and varying interactions, e.g., opening a chest with a key or throwing a coin into a piggy bank. The second task eliminates these diverse interactions. Instead, it is just a simple carrying task. Players must carry a coin in counter-clockwise rotation from one interaction point to the other. Whereas this task resembles the first one regarding movement patterns, it is deliberately designed to feel utterly annoying.

\subsection{Procedure and Applied Measures}
We conducted a user study with a mixed design. All participants were randomly split into four groups, each playing both tasks with one of the four wall designs. The study was executed in our VR lab using an HTC Vive Pro~\cite{vive}. We started by informing the subjects about the study process without giving away our research focus. Subsequently, the participants completed a general questionnaire assessing gender, age, and prior VR and gaming experiences. Finally, we administered the Immersive Tendencies Questionnaire (ITQ)~\cite{Witmer.1998} to assess the participants' tendency to immerse in fiction.

After we introduced the participants to our VR equipment, they played the first round. Upon completion, the subjects removed the HMD and filled out the IGroup Presence Questionnaire (IPQ)~\cite{Schubert.1999b}. We administered the IPQ to assess whether the wall designs influence the perceived presence, serving as an explanatory factor for H2. Subsequently, the subjects returned to the virtual environment to complete the second task. After finishing, the study was closed with a semistructured interview session to gather the personal reasons for ignoring or avoiding walls. Additionally, we logged the relevant playing statistics, such as the players' speed and walking distance, to confirm comparability between the two conditions. Finally, we recorded the timing and number of wall collisions. As participants might accidentally touch walls without passing them, we only counted collisions where the headset moved entirely through the wall.

\section{Results}
In total, 40 persons (20 female, 20 male) participated in our study with a mean age of $24.1$ ($SD=2.68$). The subjects equally split among nonplayers, occasional players, and regular gamers, and a majority (75\%) of them had already used VR headsets before. For the between-subject distinction, we randomly split the participants into four groups (A30: 11, A60: 9, RTH: 10, RWW: 10). Further, we did not find any significant differences regarding age, gender, prior VR experience, or immersive tendencies according to the ITQ (all $p>.05$). 

\begin{table*}[t]
  \caption{Mean scores, standard deviations, and one-way ANOVA values of the IGroup Presence Questionnaire (IPQ) for both tasks.}
  \label{tab:quest}
  \begin{tabularx}{2.045\columnwidth}
  {>{\raggedright\arraybackslash}p{3.5cm}
  >{\centering\arraybackslash}X
  >{\centering\arraybackslash}X
  >{\centering\arraybackslash}X
  >{\centering\arraybackslash}X
  >{\centering\arraybackslash}p{1cm}
  >{\centering\arraybackslash}p{0.8cm}
  >{\centering\arraybackslash}p{0.6cm}
  >{\raggedright\arraybackslash}p{0.3cm}
  }
    \toprule
     & A30 ($N = 11$) & A60 ($N = 9$) & RTH ($N = 10$) & RWW ($N=10$) & F(3,36) & $\hat{\omega}^{2}$ & $p$ & \\
    \midrule
    IPQ (scale: 0 - 6)\\
    \ \ \ \ \ \ Spatial Presence & 4.02 (1.09) & 3.91 (0.66) & 4.68 (0.92) & 4.90 (1.10) & 2.496 & .101 & .075 & \\
    \ \ \ \ \ \ Involvement & 3.07 (1.31) & 3.33 (0.77) & 4.10 (0.90) & 4.53 (1.37) & 3.631 & .165 & .022 & *\\
    \ \ \ \ \ \ Realism & 3.11 (0.70) & 2.42 (0.47) & 4.20 (0.77) & 4.03 (0.64) & 15.120 & .514 & .001 & **\\
    \ \ \ \ \ \ General & 4.27 (1.27) & 3.67 (1.12) & 5.00 (1.05) & 5.10 (0.74) & 3.766 & .172 & .019 & *\\
    \bottomrule
     &&&&&& \multicolumn{3}{c}{*\textit{p} $<.05$, ** \textit{p} $<.01$}\\
\end{tabularx}
\end{table*}

\begin{table*}[t]
  \caption{Mean scores, standard deviations, and one-way ANOVA values of the logged walking distances, average walking speeds, and average number of wall collisions for both tasks.}
  \label{tab:quest2}
  \begin{tabularx}{2.045\columnwidth}
  {>{\raggedright\arraybackslash}p{3.5cm}
  >{\centering\arraybackslash}X
  >{\centering\arraybackslash}X
  >{\centering\arraybackslash}X
  >{\centering\arraybackslash}X
  >{\centering\arraybackslash}p{1cm}
  >{\centering\arraybackslash}p{0.8cm}
  >{\centering\arraybackslash}p{0.6cm}
  >{\raggedright\arraybackslash}p{0.3cm}
  }
    \toprule
     & A30 ($N = 11$) & A60 ($N = 9$) & RTH ($N = 10$) & RWW ($N=10$) & F(3,36) & $\hat{\omega}^{2}$ & $p$ & \\
    \midrule
    Task 1: Puzzles\\
    \ \ \ \ \ \ Collisions & 0.18 (0.40) & 0.11 (0.33) & 0.00 (0.00) & 0.10 (0.32) & 0.612 & -.003 & .612 & \\
    \ \ \ \ \ \ Walked Distance ($m$) & 51.75 (14.46) & 49.11 (15.85) & 49.20 (10.70) & 53.27 (10.36) & 0.240 & -.060 & .868 & \\
     \ \ \ \ \ \ Walking Speed ($m/s$) & 0.260 (0.052) & 0.279 (0.041) & 0.254 (0.021) & 0.264 (0.045) & 0.220 & -.076 & .882 & \\
     \\
    Task 2: Moving Coins\\
    \ \ \ \ \ \ Collisions & 11.91 (17.77) & 20.11 (20.31) & 10.80 (15.92) & 6.00 (12.79) & 1.136 & .010 & .348 & \\
    \ \ \ \ \ \ Walked Distance ($m$) & 161.58 (29.77) & 172.92 (30.15) & 177.23 (42.17) & 184.14 (23.44) & 1.589 & .042 & .209 & \\
     \ \ \ \ \ \ Walking Speed ($m/s$) & 0.561 (0.078) & 0.536 (0.095) & 0.555 (0.105) & 0.545 (0.063) & 0.162 & -.061 & .921 & \\
    
    \bottomrule
     &&&&&& \multicolumn{3}{c}{*\textit{p} $<.05$, ** \textit{p} $<.01$}\\
\end{tabularx}
\end{table*}

To compare the four independent groups, we performed one-way analyses of variances (ANOVA) for the IPQ measures and most of the logged data. To meet the requirements, we ensured normal distribution with Kolmogorov-Smirnov and homogeneity of variances with Levene's tests. All measures met both conditions. This result allowed us to use Tukey's tests for all post hoc comparisons. For dichotomous data, i.e., assessing whether subjects ignored virtual walls, we used chi-squared tests of independence for comparisons between conditions and McNemar's test for comparisons between the two tasks.

\subsection{IPQ}
To determine whether our different wall designs affected the players' feeling of presence and realism, we assessed all subscales of the IPQ questionnaire. The results are depicted in Table~\ref{tab:quest}. For the two subdimensions involvement ($p=.022$) and realism ($p<.001$), as well as for the general feeling of presence ($p=.019$), the ANOVA indicates a significant difference. Post hoc comparisons indicate that the RTH ($p=.047;$ $95\%$ $CI[.011,2.655]$) and RWW ($p=.029;$ $95\%$ $CI[.111,2.755]$) conditions led to a significantly higher general presence compared to the more opaque abstract wall (A60). 

Furthermore, both realistic wall designs RTH and RWW provided a significantly higher perceived realism compared to the two abstract conditions A30 and A60, according to post hoc comparisons (A30/RTH: $p=.003;$ $95\%$ $CI[.301,1.864]$, A60/RTH: $p=.001;$ $95\%$ $CI[.966,2.601]$, A30/RWW: $p=.016;$ $95\%$ $CI[.134,1.689]$, A60/RWW: $p=.001;$ $95\%$ $CI[.791,2.426]$). For the involvement subscale, only the difference between the A30 and RWW conditions is significant ($p=.027;$ $95\%$ $CI[.127,2.787]$).

\subsection{Logged Data}
Apart from assessing the IPQ, we also analyzed the participants' play sessions by logging the individual walking trajectory, total walking distance, average walking speed, and wall collisions. The walking distance and walking speed measures did not reveal any significant differences between the puzzle task's four conditions. Similarly, the differences for both values were not significant for the repetitive task either. These results are depicted in Table~\ref{tab:quest2}.

\begin{figure*}[t!]
\centering
\includegraphics[width=2.045\columnwidth]{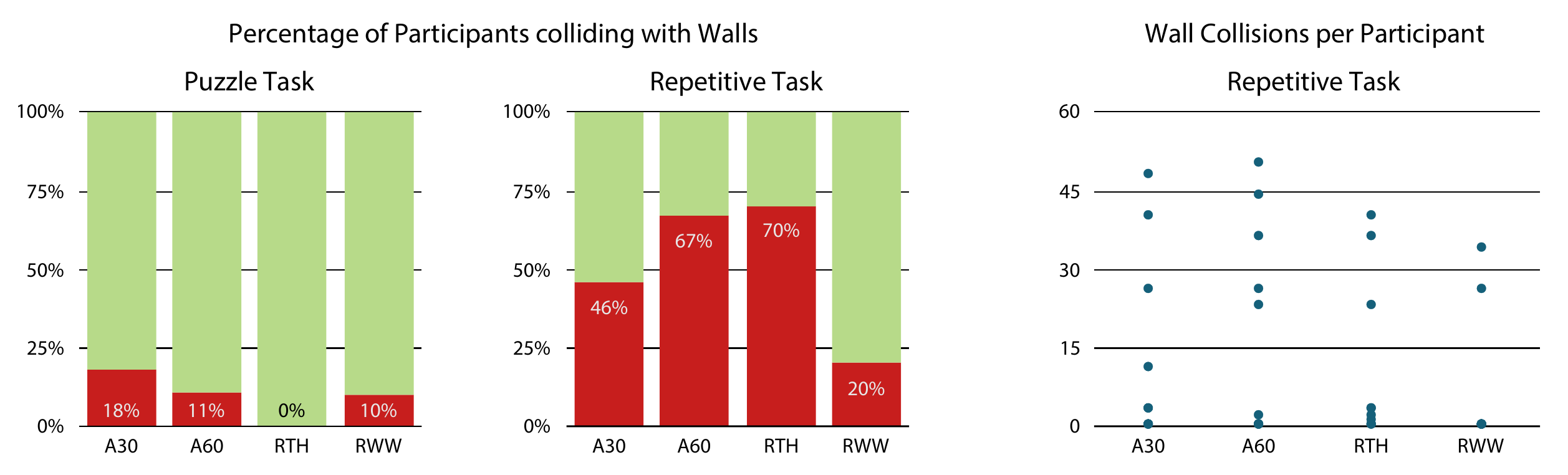}
\caption{Results from the data logged during the play sessions. Left: the percentage of subjects walking through at least one wall for each task and condition. Right: dot-plot of the wall collisions per participant and condition in the repetitive task.}
\label{fig:loggeddata}
\end{figure*}

\begin{figure}[t!]
\centering
\includegraphics[width=\columnwidth]{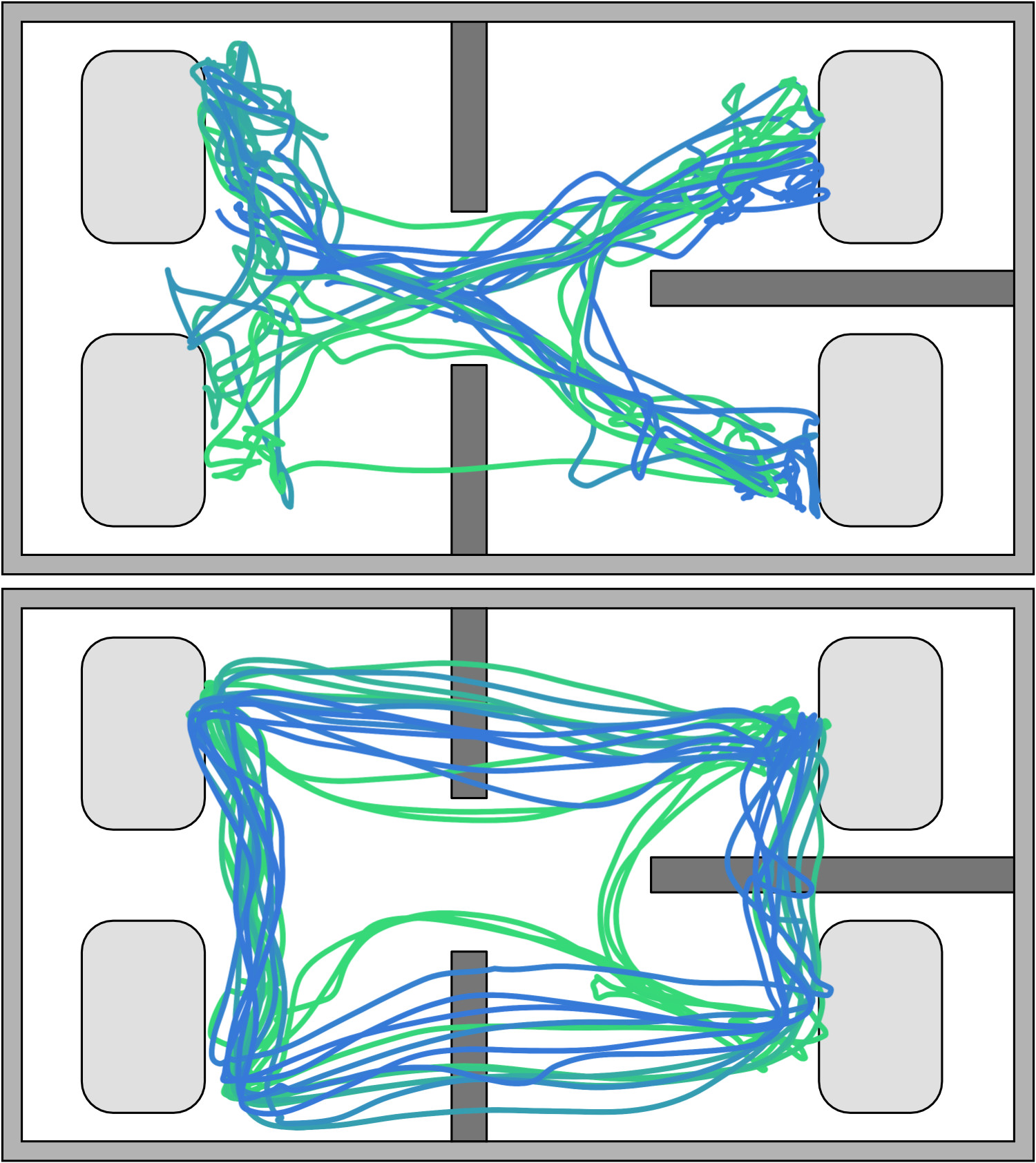}
\caption{Two exemplary walking trajectories of our study. The path's color changes from blue to green. Top: subject solving the riddle task and colliding once with a wall. Bottom: subject playing the repetitive task and ignoring the virtual walls most of the time.}
\label{fig:walking}
\end{figure}

The most important data for our analysis are the participants' behavior concerning the virtual walls (see Figure~\ref{fig:loggeddata}). While playing the puzzle task, only four subjects ($10\%$) ignored a single obstacle, whereas the others did not collide once. In contrast, 21 participants ($52.5\%$) walked through walls in the second, repetitive task. This difference is significant according to McNemar's test ($p<.0001$). Exemplary walking trajectories of both rounds are shown in Figure~\ref{fig:walking}.

When comparing the wall collisions between the four conditions for the puzzle task, we did not find any significant differences, neither between opaque and transparent walls ($\chi^2(1)=.000, p=1.000$) nor between realistic and abstract wall designs ($\chi^2(1)=1.111, p=.605$). However, for the repetitive task, the player behavior differs significantly between conditions of different opacity. Whereas 80\% of subjects in the RWW condition avoided walking through walls, this was the case in only 36.7\% of the other three groups featuring see-through walls. This difference is significant ($\chi^2(1)=5.647, p=.028$). 

On the other hand, the degree of realism had no significant influence on the behavior in the repetitive task. In the two abstract conditions A30 and A60, 60\% of subjects walked through walls, whereas in the two other conditions, 45\% ignored obstacles ($\chi^2(1)=.902, p=.527$). It is worth mentioning that -- in contrast to the puzzle task -- subjects tended to use shortcuts frequently after crossing walls once. The participants split almost exclusively into two groups: 65\% of subjects collided less than four times, whereas 35\% collided 23-50 times.

\section{Discussion}
Virtual scenarios can reach their full potential, be it for entertainment or educational purposes, only if users adhere to the environment's fundamental laws. Moving through purely virtual obstacles that do not have a physical counterpart not only harms the feeling of \textit{being there} but might also cause unwanted experiences or even result in dangerous situations. In our study, we addressed three characteristics of virtual scenarios that might foster such behavior. 

\subsection*{H1: Repetitive and monotonous tasks provoke significantly more participants to walk through virtual walls than diverse tasks.}
We approached this hypothesis by integrating two similar tasks in a within-subject design into our study. Both required the players to move virtual items between interaction points spread across the play area. The necessary completion time and walking distance were chosen similarly so that the task's motivation remained the only variation. The significant difference between both tasks regarding players that cut short through walls, i.e., $10\%$ versus $52.5\%$, confirms our hypothesis.

Furthermore, we observed that all subjects, who walked through walls in the diverse puzzle task, tried this shortcut only once. In contrast, 35\% of the participants ignored most of the walls in the repetitive condition. This finding further supports our initial assumption that player behavior mainly relies on personal interest in the situation. Varying and interesting assignments preserve the scenario's plausibility and provide a solid incentive to stick to the rules. On the other hand, repeated and simple actions fail to keep the players immersed in the virtual world. As oral feedback suggests, participants were more aware of the real situation and looked forward to completing the task: \textit{"I knew that the walls were not real, so I just walked through them to get this annoying task done as fast as possible."(\textbf{P14})}.

\subsection*{H2: Participants walk significantly more often through abstract walls than through realistic walls.}

Among the four walls that we tested in a between-subject design, two were implemented as abstract blocks (A30 \& A60), and the other two thematically fit the testbed scenario (RTH \& RWW). We assumed designs that matched less would serve as a \textit{break-in-presence}~\cite{Slater2000Presence}, diminishing the visual realism and consequently leading to less conform behavior. By administering the IPQ, we confirmed that the abstract walls indeed harmed the perceived presence and realism. Additionally, subjects later reported that the walls in these conditions \textit{"felt like unfinished placeholders and somewhat ruined the appealing visuals of the scenery"(\textbf{P2})}. 

However, these observations did not affect the participants' walking behavior. Even though the percentage of subjects walking through walls was tendentially higher for the abstract conditions (in the repetitive task: 56.1\% versus 45\%), this difference was not significant. Therefore, we cannot confirm our second hypothesis. Considering the observed tendencies in the data, we suspect the potential presence of a less noticeable effect that was overlain by the strong findings of H1 and H3. Thus, we propose further research to investigate this open question in isolation.

\subsection*{H3: Opaque surfaces deter more participants from ignoring virtual obstacles than partly transparent surfaces.}

Apart from varying the virtual walls in the degree of realism, we also used multiple opacity levels. For this hypothesis, we group A30, A60, and RTH into one category of see-through designs. Even though the twine material was fully opaque, the underlying wall model featured numerous holes, clearly revealing the other side. In contrast to these designs, the wooden surface completely blocked the view of objects behind the wall. This differentiation between conditions resulted in significantly different behavior observed in the repetitive task: 63.3\% of the subjects in the see-through conditions deliberately ignored walls, compared to only 20\% in the RWW condition. 

Subjects often reported the wall's transparency as assuring factor in their decision-making: \textit{"I saw that my goal was right behind the wall. Since I knew that there were no free-standing walls in the room, I felt safe to walk through."(\textbf{P8})}. Similarly, participants in the RWW group felt deterred by the solid appearance of the wall: \textit{"Of course I knew that these walls were only virtual. But they appeared so sturdy that I preferred to walk around."(\textbf{P11})}. The oral feedback shows that users generally prefer the safety of seeing where they are going and refrain from walking into unclear areas. Together with the recorded data, this finding confirms our hypothesis that opaque surfaces deter players from walking through walls.

\subsection*{RQ1: How do players decide whether they pass through virtual walls or walk around them?}

Apart from investigating our three main hypotheses, we were also interested in the participants' reasons for deciding whether they walk through or around virtual walls. Thus, we followed the main study with a semistructured interview allowing the subjects to share their personal thoughts. We analyzed the resulting interview data for reoccurring motives using a peer-reviewed deductive thematic analysis~\cite{braun2006thematicAnalysis} and structured the reasons into two categories.

\subsection{Reasons for refraining from walking through obstacles}

The overwhelming majority of subjects stated a simple reason for not even considering walking through obviously virtual walls: \textit{"Walls are solid, you cannot walk through them."(\textbf{P3})}. This feedback indicates a strong \textit{Plausibility Illusion}. The participants transferred the real world's fundamental rules to the virtual scenario and stuck to basic physical principles, treating the virtual environment like its material counterpart.

Many subjects also reported their fear of negative consequences when breaking the rules. This reason encloses a variety of partly subconscious considerations. Some participants felt unsure not being able to see behind the wall: \textit{"I would not have seen what was directly in front of me, so I decided to stay cautious."(\textbf{P16})}. Others feared hurting themselves: \textit{"The twines seemed painful -- I usually avoid touching such hedges."(\textbf{P23})}. Finally, subjects also expected to get punished for nonadherent behavior, such as \textit{"having to restart the level"(\textbf{P20})}.

\subsection{Reasons for walking through obstacles}

The most commonly mentioned reason for ignoring the walls in the puzzle task was curiosity. Participants were eager to explore their abilities in the virtual world and test the game's rules. However, after experiencing the absence of any punishment, all subjects reverted to an adherent playstyle for the remaining time: \textit{"It was interesting to revolt against the intended playstyle. But after trying once, I decided that walking around the walls was more fun and felt more natural."(\textbf{P1})}.

For the repetitive task, most players reported a different reason for their behavior: cutting short. As this round consisted only of the always same interaction of putting coins on matching plates, subjects mostly decided to act pragmatically and \textit{"choose the shorted possible path, even though walking through the walls felt awkward and unnatural"(\textbf{P19})}. The logged walking trajectories support this feedback as many participants began walking as intended and only started cutting short after realizing that the task would not change for the remainder of the round.

Finally, few subjects in the more transparent A30 condition mentioned a different reason not observed for any other group. These players did not recognize the walls as obstacles being part of the virtual environment. Instead, they had the impression that \textit{"these vitreous-looking cuboids must have been some kind of graphical artifact that had no particular meaning. It did not match the underwater scenario and was barely visible, so I thought it was safe to ignore."(\textbf{P26})}. This feedback is in line with findings by Simeone et al.~\cite{Simeone2017}, who reported that incorrect interpretations could lead to arbitrary behavior.

\section{Conclusion and Future Work}

When playing VR games, users might not always behave in the intended way or adhere to the fundamental principles of virtual scenarios. While developers can prevent most of such behavior through prescient game design, some challenges remain. One prominent example is the discrepancy between the physical surroundings and the virtual world. When using real walking, players can walk freely within the play area's boundaries, despite virtual obstacles blocking their path. It is not possible to prohibit such behavior without losing the valuable advantages of natural walking. Therefore, past research has mainly concentrated on providing multisensory collision feedback to deter players from deliberately walking through virtual walls. However, it remained unclear how the virtual scenarios' characteristics, such as wall design or task type, influence the players' incentives toward this behavior.

In our work, we investigated how players reacted when confronted with different situations. We thus concentrated on three potentially influencing factors: task type, wall opacity, and wall realism. In a mixed study, we confronted the subjects with two iterations of item-carry tasks. One round featured an engaging motivation, whereas the other lacked any interesting variety and focussed on repetition only. Participants in the repetitive condition cut short significantly more often to finish the assignment as fast as possible. Further, we varied the tested wall layouts in a between-subject design to cover different degrees of realism and opacity. Even though significantly more subjects ignored transparent walls than solid ones, we could not find a similar difference between realistic and abstract designs. 

Our findings reveal that various factors influence the individual decision whether players adhere to or ignore obstacles in the virtual scenario. One of the most decisive parameters is the task type. Challenges that keep the players engaged and interested in the experience are particularly effective in avoiding nonadherent behavior. In contrast, repetitions, simple interactions, or forced obligatory tasks diminish the users' feeling of presence and foster pragmatic actions. In such cases, opaque surfaces deter more players from walking through virtual walls as these hide the destination and cause doubts regarding the penetrability. These findings extend the existing research corpus on player behavior and help developers design virtual scenarios that reach their full potential.

Future studies are needed to investigate the open questions on player behavior in the context of virtual collisions. Even though we did not find any evidence for an influence of the degree of realism, we suspect that our other findings might have concealed a potential minor effect. Apart from this lack of clarity, we also aim to research whether the observed behavior might be affected by individual player characteristics. Throughout the study, we noticed that participants either walked strictly around every wall or always took the direct path. However, we did not arrive at a final explanation for this almost binary classification. Finally, how time pressure might alter the observed behavior remains to be investigated. Strictly timed tasks were often used by prior research to provide a strong wall-ignoring incentive. However, this approach was based upon personal experiences and has not yet been observed in isolation.

\bibliographystyle{IEEEtran}
\bibliography{IEEEabrv,literature}

\end{document}